\documentclass[aps,prb,reprint,floatfix,amsmath,amssymb,
superscriptaddress,nofootinbib]{revtex4-1}

\usepackage{graphicx}
\usepackage{epstopdf}
\usepackage{color}
\usepackage{printlen}
\usepackage{units}
\usepackage{booktabs}
\usepackage{physics}

\renewcommand{\vec}[1]{\mathbf{#1}}

\renewcommand{\Im}{\text{Im}}

\PassOptionsToPackage{unicode}{hyperref}
\PassOptionsToPackage{naturalnames}{hyperref}

\begin{document}

\title{Unconventional Flat Chern Bands and  2$e$ Charges   
in  Skyrmionic  Moir\'{e} Superlattices}

\author{Yifei Guan}
\affiliation{Institute of Physics, Ecole Polytechnique F\'{e}d\'{e}rale de Lausanne (EPFL), CH-1015 Lausanne, Switzerland}      
\author{Oleg V. Yazyev}
\affiliation{Institute of Physics, Ecole Polytechnique F\'{e}d\'{e}rale de Lausanne (EPFL), CH-1015 Lausanne, Switzerland}    
\author{Alexander Kruchkov} 
\affiliation{Institute of Physics, Ecole Polytechnique F\'{e}d\'{e}rale de Lausanne (EPFL), CH-1015 Lausanne, Switzerland}    
\affiliation{Department of Physics, Harvard University, Cambridge, MA 02138, USA}
\affiliation{Branco Weiss Society in Science, ETH Z\"urich, CH-8092 Z\"urich, Switzerland}
\date{\today}

\begin{abstract}
The interplay of topological characteristics in real space and reciprocal space can lead to the emergence of unconventional topological phases.  
In this Letter, we implement a novel mechanism for generating higher-Chern flat bands on the basis of twisted bilayer graphene (TBG) coupled to topological magnetic structures in the form of the skyrmion lattice.  
In particular, we discover  a scenario for generating $|C|=2$ dispersionless electronic bands when the skyrmion periodicity and the moir\'e periodicity are matched. 
Following the Wilczek argument, the statistics of the charge-carrying excitations in this case is \textit{bosonic}, characterized by electronic charge  $Q =2e$, that is \textit{even} in units of electron charge $e$.  
The required skyrmion coupling strength triggering the topological phase transition is realistic, with its threshold estimated as low as 4~meV. 
The Hofstadter butterfly spectrum of this phase is different resulting in
an unexpected quantum Hall conductance sequence $\pm  \frac{2  e^2}{h},  \ \pm  \frac{4 e^2}{h}, ...$  for TBG with 
skyrmion order. 
\end{abstract}

 \maketitle

The interplay between the real- and momentum-space topologies is a new direction in exploring interacting topological phases of matter.  Generically, the band topology was introduced in condensed matter physics through the quantum-Hall-like arguments,\cite{TKNN,Haldane1988} which are typically insensitive to the real-space  defects. Nevertheless,  it later became clear in the context of quantum Hall ferromagnetism that the coexistence of real-space topology (skyrmions) and momentum-space topology (Chern numbers) is possible for the flat bands.\cite{Sondhi1993,Fertig1994} More broadly, the question can be formulated in the form: Can the real space topology influence the change in the momentum space topology? The answer is positive;\cite{Lux2021} and we provide a concrete example of a realistic system.

The skyrmions have been  called upon to explain possible pairing mechanisms in twisted bilayer graphene (TBG) and similar heterostructures, \cite{Khalaf2021,Chatterjee2020} see also Refs.~\onlinecite{Zhang2019,Abanov2001a,Grover2008}. In particular, the skyrmion-based theories are able to address not only the problem of possible mechanism for unconventional superconductivity in TBG,\cite{Cao2018} but also provide reasonable explanation\cite{Chatterjee2020} for nonmonotonic magnetoresistance observed in experiments.\cite{Sharpe2019,Serlin2020} Thus, the coexistence of the skyrmion order and the moir{\'e} potentials, to one or another extent, seems feasible. Since this is a consistent theoretical mechanism, we ask the following question: What is the effect of the skyrmion order on the TBG flat bands themselves?

The topological flat bands can induce a non-collinear magnetic order, such as skyrmions.\cite{Fertig1997} In general, however, the skyrmion excitations are neither bosonic, nor fermionic, and reshape themselves continuously between the two opposite quantum statistics.  \cite{Wilczek1983} It can be shown that excess \textit{electric charge} density $\rho (\vec r)$, polarized by the presence of a skyrmion $\vec S (\vec r)$ on a flat Chern band is given by \cite{Sondhi1993,Abanov2001a,Hsu2013} 
\begin{align} 
\rho (\vec r) = C e \, \frac{\vec S \cdot  (\partial_x \vec S  \times   \partial_y \vec S ) }{4 \pi}, 
\label{density-r}
\end{align}
\noindent
where $C$ is the Chern number of the underlying electronic band. Upon integration over the unit cell one obtains 
\begin{align}
Q= C \mathcal{W} e, 
\label{charge}
\end{align}
where $\mathcal W$ is the skyrmion winding number (an integer). Thus,  skyrmions polarize a discrete electric charge when the underlying flat band is topological.

The collective effect of the real-space and momentum-space topologies in mechanism \eqref{charge} plays an important role. Indeed, the effective magnetic field, assigned to the skyrmion $B (\vec r) \propto \vec S \cdot  (\partial_x \vec S  \times   \partial_y \vec S )$  in \eqref{density-r} is essentially the Berry field.\cite{Nagaosa2013} In momentum space, the gluing mechanism \eqref{charge}  can be rewritten as
\begin{align}
\rho (\vec q) =e \sum_{\vec k} c^{\dag}_{\vec k +\vec q} c^{}_{\vec k} =  \mathcal{W} \frac{F_{xy}}{2\pi},  
\label{density-q}
\end{align}
\noindent 
where $F_{xy}$ is the Berry curvature associated with the Chern band, $C = \frac{1}{2 \pi} \int_{\text{BZ}} F_{xy} dk_x dk_y$. Thus, according to Eq.~\eqref{density-q}, it is possible to have an unconventional electronic pairing mechanism induced by the skyrmion order when electrons are moving in the Chern band. This has been recently pointed out in the models of skyrmionic superconductivity in TBG, \cite{Chatterjee2020,Khalaf2021} where  unconventional pairing mechanism is resulting from the skyrmions in the pseudospin space built upon reshuffling intrinsic $C=+1$ and $C=-1$ sectors.\cite{Khalaf2021}  
However, a  different situation is possible if the moir{\'e} system develops a flat band of a \textit{higher} Chern number. Namely, in the $C=2N$ case, the electric charge polarized by the skyrmion is 
\begin{align}
Q = 2 N \mathcal W e,  
\label{charge-2e} 
\end{align}
that is {\it bosonic}.
In other words, in case of $|C|=2$ electron pairs are preformed independently of mechanism developed in Ref.\cite{Khalaf2021}.
In this Letter, we show that the  flat bands in TBG acquire even Chern numbers $|C|=2$ when electrons are firmly coupled to the underlying skyrmion lattice.

\begin{figure*}[t]
\includegraphics[width = 0.8 \textwidth]{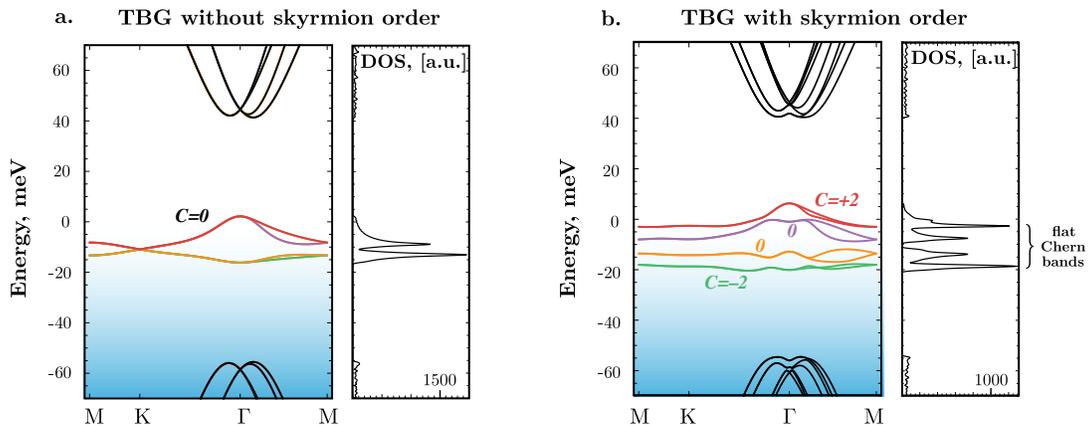}
\caption{\textbf{{Unconventional Chern flat bands of $|C|$$=$$2$} in twisted bilayer graphene matched with the skyrmion lattice.} (a) Band structure and density of states (DOS) of magic-angle TBG without skyrmion order. The associated Chern number is zero. (b) The inclusion skyrmion order ($m_0=4$~meV, $\mu=8$~meV) results in flat Chern bands with $|C|=2$. }
\label{bandstructure}
\end{figure*}

\textbf{Underlying skyrmion lattice}.---The calculations below are independent of a particular origin of the skyrmion lattice (SkL), and can be applied to both (i) the case of proximity to a substrate hosting skyrmion lattice, or (ii) the case of spontaneous formation of skyrmion lattice in the moir\'e system. 
In the former scenario (i), the periodicity of SkL $\lambda_{\text{SkL}}=J/2D$ dictated by the interplay of Heisenberg ($J$) and Dzialoshinsky-Moriya ($D$) interactions can be engineered\cite{Tokura2020} to match the periodicity of the moir\'e superlattice of TBG ($a_\text{M} = 12.9$~nm at the magic angle $\theta=1.08^\circ$).
The later scenario (ii), represented by the SkL forming as a result of  instability associated with the flat band, is supported by recent theoretical calculations.\cite{Wu2020,Bomerich2020}  At $\nu=3/4$ filling of the conduction band, TBG has been reported to exhibit giant spontaneous magnetism, resulting from internal Berry fields of the band structure, that is orbital in nature.\cite{Sharpe2019, Serlin2020}  It is interesting that the TBG flat band can be viewed as a Landau level in non-homogeneous magnetic field,\cite{TKV,Popov2021,SanJose2012} and this real-space Berry field has strong inhomogeneities of the same periodicity as the patern moir{\'e} potential.\cite{Ledwith2020}  Thus, the resulting SkL instability, if allowed by the mechanism of Ref.~\onlinecite{Wu2020}, will naturally inherit the same periodicity as the pristine TBG moir\'e superlattice.
In what follows below, we consider the skyrmion lattice matched with the periodicity of TBG, but similar principles can be applied to other moir{\'e} heterostructures.

\begin{figure*}[t]
	\centering
	\includegraphics[ width = 0.85 \textwidth]{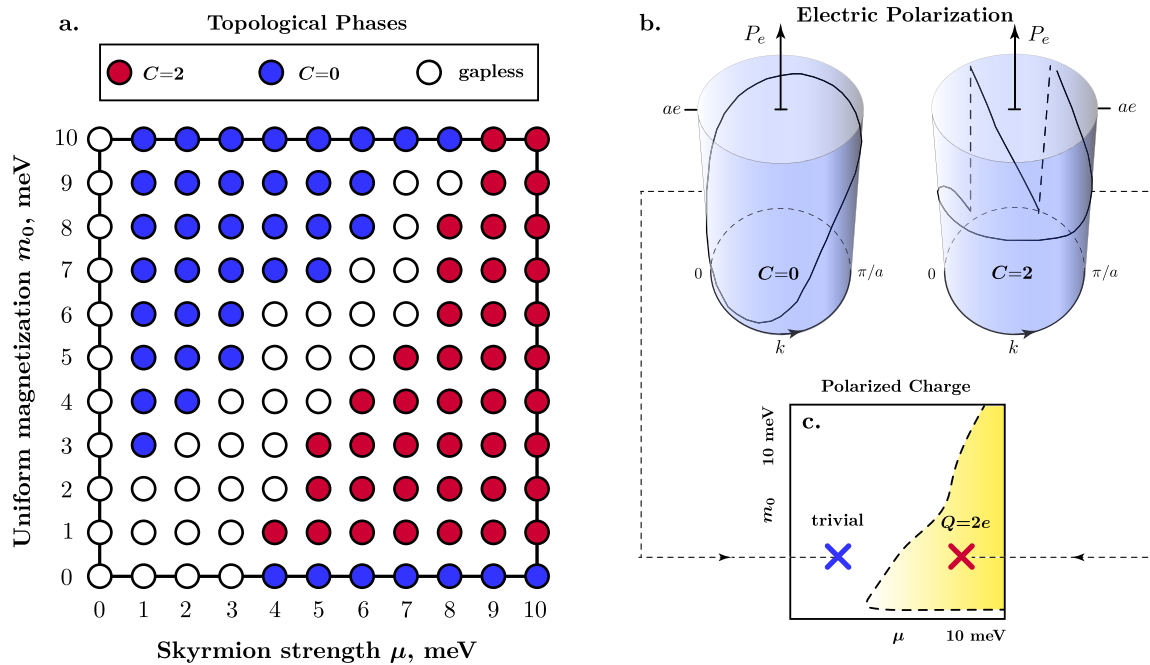}
	\caption{{(a)Phase diagram of flat Chern band phases in twisted bilayer graphene with the skyrmion order.} We see that at the low skyrmion field the topologically--trivial phases are energetically preferred. However, at higher skyrmion fields, the phase diagram shows a remarkable new phase of $C$$=$$2$ flat composite bands.
{(b) The calculated electric polarization of the different TBG phases with skyrmion lattice, see Eq. \eqref{polarization}. The numerical parameters used: $\mu=8$ meV, $m_0 = 4$ mev ($C=2$ cylinder) and $\mu=2$ meV, $m_0 = 4$ mev ($C=0$ cylinder).   (c) Sketched phase volume depicting the polarized charge 2$e$ (in units of Fig.2a). } 
}
	\label{phasediagram}
\end{figure*}

\textbf{Band structure and unconventional Chern flat bands.}---We start by addressing {the effect of skyrmion order} on the band structure of TBG. 
We  demonstrate that the skyrmion lattice, when being \textit{commensurate} with the moir\'e periodicity $a_{\text{M}}$ at the first magic angle, changes the electronic response of the system beyond recognition. We define the underlying SkL as \cite{Muhlbauer2009,Nagaosa2013} 
\begin{align}
\vec m (\vec r) = \vec m_0 + \mu \sum^3_{j=1} (\vec S_{\vec b_j} e^{i \vec b_j \vec r} + \vec S_{-\vec b_j} e^{-i \vec b_j \vec r} ),
\label{SkL}
\end{align}
where $m_0$ is the uniform magnetization component, $\mu$ the SkL strength and reciprocal vectors  $\vec b_1 = \vec q_2 - \vec q_1$, $\vec b_2 = \vec q_3 - \vec q_1$, $\vec b_3 = \vec q_3 - \vec q_2$ are the same for the SkL and the moir\'e superlattice (here $\vec q_1 = k_{\theta}(0,-1)$, $\vec q_{2,3} = k_{\theta} (\pm \sqrt{3}/2, 1/2)$, with moir{\'e} wave number $k_{\theta} = 2k_D \sin \frac{\theta}{2}$, $k_D$ being Dirac momentum of the monolayer graphene and $\theta$ being the twist angle).
For the considered ranges of parameters $m_0,\mu = 0..10$~meV, the effective local magnetic fields are sufficient to modify the band structure (Fig.~\ref{bandstructure}) and promote unconventional Chern phases (Fig.~\ref{phasediagram}).

We focus our attention on the effect of the underlying skyrmion order \eqref{SkL} on the electronic band structure of TBG at the magic angle $\theta = 1.08^{\circ}$. 
We have first confirmed this effect within the continuum model (see Supplemental Materials), and then performed the tight-binding model calculations for spinful electrons on a magic-angle TBG configuration that takes into account lattice relaxation effects.
The effect of skyrmion spin texture is introduced via the on-site terms
$H_{s}(\vec r_i)=\boldsymbol{\sigma} \cdot \vec m(\vec r_i)$ for each of the orbitals.
Further details of our calculations are presented in the Methods section.

Figure~\ref{bandstructure}a compares the electronic band structure of TBG without the skyrmion order with that (Fig.~\ref{bandstructure}b) for a representative case characterized by $m_0=4$~meV and $\mu=8$~meV.
Importantly, these parameters result in the $Q=2e$ phase as shown in phase diagram in Fig.~\ref{phasediagram}a.  
In absence of SkL ($m_0=0$~meV, $\mu=0$~meV) magic-angle TBG has 8 flat bands near the Fermi level isolated from remote bands (Fig.\ref{bandstructure}a). 
At the $K$ point of mini Brillouin zone, the flatband manifold has degenerate Dirac-like band crossing with renormalized Fermi velocity, while the degeneracy at the $\Gamma$ point due to the spin and valley degrees of freedom.
Introducing the uniform magnetization ($m_0 \ne 0$) lifts the spin degeneracy splitting the flat band manifold into 2 groups, while leaving intact the degeneracy at the Dirac points {(see the Supplemental Material)}. 
In the presence of skyrmion order ($\mu \ne 0$), the spin-orbit {terms} couple the spin with the bands opening gaps at points $K$ and $K'$ and contributing to the hybridization of flat bands at the $\Gamma$ point (Fig.~\ref{bandstructure}b). 
In this case, the 8 flat bands decompose into 4 doubly degenerate sub-bands, which remain significantly flat. Employing the band flatness criterion of Bistritzer and MacDonald\cite{Bistritzer2011} (renormalized Fermi velocity $v_F$ the at the $K$ point), we observe that these bands become very flat in the $K$, $K'$ valleys ($v_F=3.3\times10^4$~m/s  without SkL vs. $v_F=2\times 10^2$~m/s with SkL). For comparison, the overall bandwidths in TBG with SkL are significantly lower than in the pristine TBG case (5.3 mev,  13 meV  in pristine TBG and    2.7 meV, 5 meV, 8.5 meV, 9.3 meV in TBG with SkL, given in ascending bandwidth order). Importantly, the density of states (DOS) remains significant upon adding SkL term (Fig.~\ref{bandstructure}), thus enabling strong correlations. 
To quantify the the band flatness further, we investigate the band flatness criterion in terms of wave functions.\cite{Kruchkov2021} 
We thus confirm that  upon adding the SkL term, the flat bands  remain well-defined with clear real-space localization resulting into significant electronic density of states (Fig.~\ref{bandstructure}b), with two of the resulting flat Chern bands having the high Chern number $|C|=2$. 
Thus, non-collinear spin texture of skyrmion lattice coupled to charge-carrier in twisted bilayer graphene modifies its original band structure, but preserves the flatness in the atomistic model with lattice relaxations  We now proceed to  the effect on band topology.

 \begin{figure*}[t]
    	\includegraphics[width= 0.85 \textwidth]{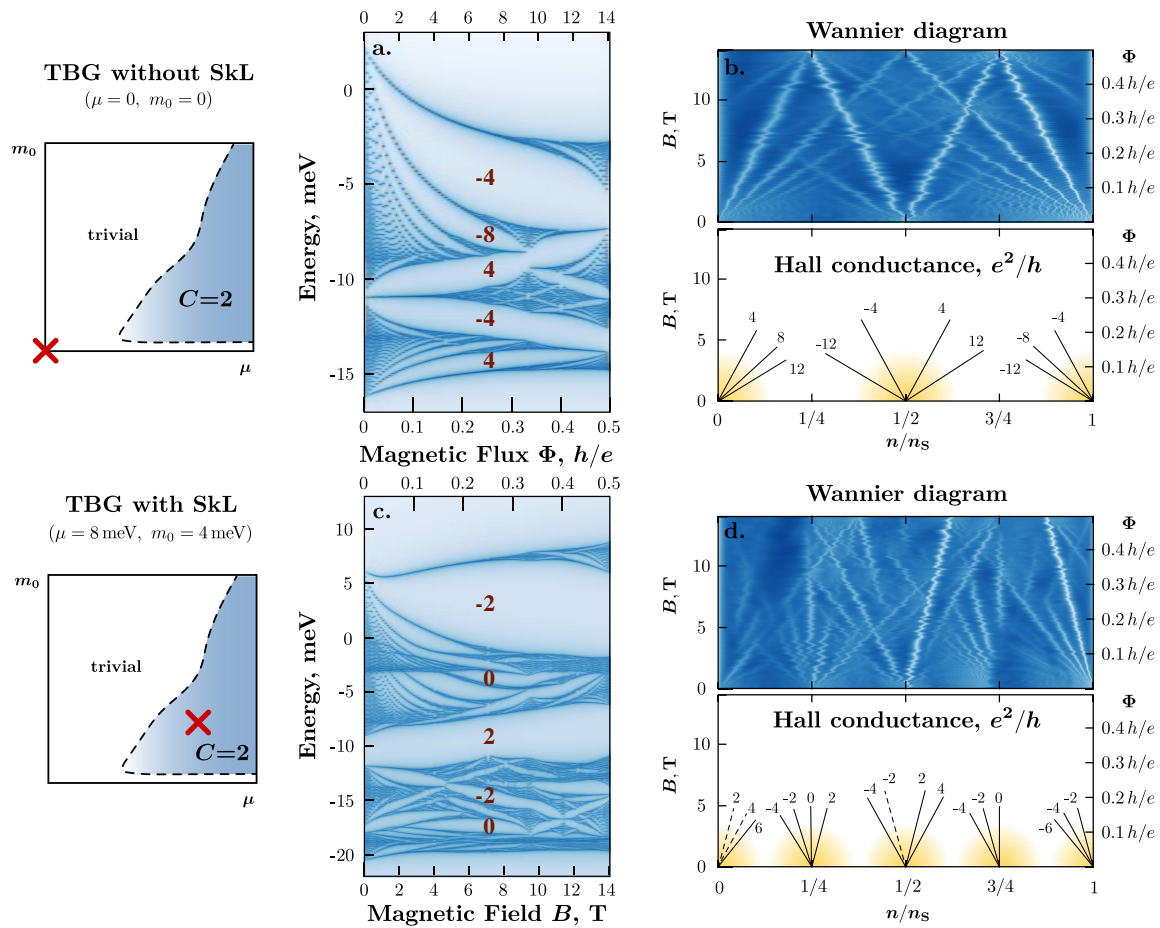}
	\caption{\textbf{{Hofstadter and Wannier diagrams of TBG without and with the skyrmion order}} Hofstadter (a) and Wannier (b) diagrams of twisted bilayer graphene without the skyrmion order.  Hofstadter (c) and Wannier (d) diagrams of twisted bilayer graphene with the skyrmion order. 
	 We clearly see a different pattern of gaps and Chern numbers, attributed to the magnetic-field responds of a flat Chern band (see main text). All calculations are performed with the tight-binding model accounting for atomic relaxations of TBG. Here $n_S =8$ (total bands) and $\Phi_0 = h/e$.}
	\label{Hofstadter}
\end{figure*}

\textbf{Skyrmion-induced band topology and the phase diagram.}--- The pristine TBG has `band structure with hidden Chern numbers $C = \pm 1$ per valley that sum up to zero Chern number in total. 
Thus, in absence of  skyrmion lattice or explicit symmetry  breaking, TBG is not expected to demonstrate Chern indices of the composite flat bands (i.e. $m_0=\mu=0$~meV in phase diagram shown in Fig.\ref{phasediagram}a is topologically trivial). 
In what follows below, we discuss the half-filling of the conduction and valence bands ($\nu = 3/4$ and $\nu = 1/4$ fillings of the 8 moir{\'e} flat bands, respectively),  motivated by the experimentally relevant filling $\nu = 3/4$ (see e.g. Ref.\cite{Sharpe2019}). 
The formalism of hybrid Wannier functions (HWFs)\cite{Soluyanov2011} is employed to calculate the positions of the Wannier charge centers (WCC). 
This formalism gives us advantage of both addressing the Chern numbers as the winding of WCC on the cylinder (see Fig.~\ref{phasediagram}b), and linking the nontrivial winding to electric polarization: The sum of all Wannier charge centers $\langle W^{h}_n | \vec r| W^{h}_n \rangle$ is gauge-invariant (mod lattice constant), and relates to the electric polarization of the system\cite{King1993,Resta1993,Coh2009}
\begin{align}
   P_e=e \sum_n \langle W^{(h)}_n | \vec r| W^{(h)}_n \rangle.
   \label{polarization}
\end{align}
Upon including the skyrmion order the  $C=2$ phase exhibits $\Delta P_e=2 e a$, witnessed through the polarized charge $2e$. 
For definiteness, in Fig.~\ref{phasediagram}a and the text below, the phase diagram traces the total Chern number {at $\nu=3/4$}, that is of the lowest three composite flat bands, {which is the same in absolute value as the Chern number of the upper composite flat band,} in  the range from 0 to 10 meV for both the uniform component $m_0$ and the skyrmion strength $\mu$. 
This range is dictated by the typical bandwidth scale in the system.
Larger values of $m_0$ (over 20 meV) drives the flat bands to overlap with remote bands and destroys the separation condition of the flat bands, while $\mu$ does not induce the overlap explicitly.\\ 

The phase diagram in Figure~\ref{phasediagram}a can be  divided into three parts: two topologically-trivial phases (gapped $C=0$ phase and gapless phase denoted "gapless" for which the Chern numbers cannot defined).  
Below the critical value $\mu\approx4$~meV, the system remains in one of the trivial phases. Upon increasing the skyrmion strength above $\mu = 4$~meV, the new topological phase with $C=2$ emerges. This phase, however is possible if only both $\mu$ and $m_0$ are finite: As shown in Fig.~\ref{phasediagram}a, the WCC winding is zero when $m_0=0$. 
One can find a tiny band gap near the $\Gamma$ point at $m_0=0$, in which the Chern number is zero. However, 
 a weak ferromagnetic component $m_0 \ll \mu$ is sufficient to drive the system into the $C = 2$ phase, provided the skyrmion strength  with $\mu > 4$~meV. 
On the other hand, the ferromagnetic component $m_0$ does not contribute to the finite Chern number itself, since in the region $\mu < 4$~meV  the total Chern number is always zero (or undefined as a consequence of gaplessness). We highlight that the skyrmion strength of 4 meV is a realistic value, comparably smaller than the bandwidth itself, and a small uniform magnetization turns the system into a higher-Chern topological phase.

\textbf{Hofstadter spectra}.--- 
Another qualitative difference of the system with skyrmionic order can be traced in the response to the magnetic field. In this connection, the established apparatus involves the Hofstadter butterfly spectrum, which depicts the evolution of Landau levels in the system versus magnetic flux.\cite{Hofstadter1976}  
The Hofstadter butterfly spectra, together with the Wannier diagrams,\cite{Wannier1978} provide a comprehensive and sensitive map of the Chern numbers, and is directly linked to the observable quantum Hall conductance sequences.\cite{Dean2013,Hunt2013,Ponomarenko2013} In Figure~\ref{Hofstadter}, we show the difference between the magnetic field response of magic-angle TBG with and without the skyrmion order. 
For the skyrmion-free TBG, the Landau levels at small magnetic fluxes are clearly recognizable, and their field-dependence in the limit of low field is consistent with the behavior of Dirac electrons scaling as $\sqrt{B}$ (Fig.~\ref{Hofstadter}a). In this limit, the Hofstadter spectrum is energetically bound to the bandwidth of the original flat band, which is expected for the trivial phase.\cite{Herzog2020}  
Upon increasing the magnetic flux, a moderate broadening of the Landau levels is observed, while sufficiently large gaps separate low-index Landau levels.

The situation changes dramatically upon including the skyrmion order (Fig.~\ref{Hofstadter}c; for $m_0=4$~meV and $\mu=8$~meV). First of all, the low-field behavior is no longer recognizable  as a simple $\sqrt{B}$ behavior since the Dirac physics is no longer relevant in this case. Instead, the behavior is qualitatively consistent with universal response of a Chern band to magnetic fields.\cite{Herzog2020} A different pattern of pronounced gaps emerges upon increasing the flux. The broadening and hybridization between sub-bands is promoted, leading to a more complex butterfly structure. We observe that the Hofstadter spectrum of the flat band is not bounded by the bandwidth of the original flat band, and can hybridize with other bands, which in turn trivializes the band topology leading to $C = 0$ gaps. Importantly, the particle-hole asymmetry is pronounced, which results in the particle-hole asymmetric Wannier diagrams {and asymmetric Hall response}. Since the TBG with SkL host a Chern band characterized by $C=2$, we observe a different pattern of in-gap Chern numbers, as indicated in Fig.~\ref{Hofstadter}c.  In order to draw the quantitative difference between these two cases, we proceed to the analysis of the Wannier diagrams.

\renewcommand{\arraystretch}{1.5}
\begin{table}[t]
\begin{ruledtabular}
\caption{Summary of quantum Hall conductance sequences at neutrality. \vspace{2 mm}}
\begin{tabular}{p{1.5 in} |   p{1.5 in} }
TBG ($C_3$ symmetric) &  $\pm 4 \frac{e^2}{h}$, $\pm 12 \frac{e^2}{h}$, ...   \\
\hline
TBG+SkL ($C$$=$$0$ phase) &  $\pm 4 \frac{e^2}{h}$ , $\pm 8 \frac{e^2}{h}$ , ...     \\
\hline
TBG+SkL ($C$$=$$2$ phase) &  $\pm 2 \frac{e^2}{h}$ , $\pm 4 \frac{e^2}{h}$,   ...   \\
\hline
TBG experiments &  $\pm 2 \frac{e^2}{h}$ , $\pm 4 \frac{e^2}{h}$, ... + further sample-dependent$^{40}$ \\
\end{tabular}
\end{ruledtabular}
\end{table}

\textbf{Hall conductance sequences.}---The Wannier diagram \cite{Wannier1978} is obtained  from the Hofstadter butterfly spectrum by presenting the statistical weight of the states below a given gap versus the magnetic flux. Each line in the Wannier diagram depicts a gap in the Hofstadter butterfly spectrum,  while its slope characterizes the in-gap Chern numbers {($dn/dB=C e/h$)}. In Figs.~\ref{Hofstadter}b,d we plot the Wannier diagrams that correspond to the respective Hofstadter spectra shown in Figs.~\ref{Hofstadter}a,c. 
To facilitate the analysis, we overlay the lines that trace the minimal intensities in the Wannier diagrams (Fig. 3c,3d).  We find that for pristine TBG with the preserved $C_3$ symmetry, the relevant Hall conductance sequence is $\pm 4 \frac{e^2}{h}$, $\pm 12 \frac{e^2}{h}$,  ..., that is in increments of 8 conductance quanta.\cite{Zhang2019b} On contrary, the presence of skyrmion order changes the Hall sequence dramatically. For the TBG with SkL in the $C=2$ phase, the Hall conductance "fingerprint" (Fig. 3) at neutrality reads
\begin{align}
G_{\text{TBG}}^{\text{SkL}} = \pm  \frac{2  e^2}{h},  \ \pm  \frac{4 e^2}{h},  ...  
\ \ \ 
 \end{align}
One might be tempted to interpret these results as a trivial lifting of the spin degeneracy by the uniform component $m_0$ of the skyrmion phase. However, we observe that by changing the skyrmion parameters to $m_0 = 4$~meV and $\mu= 2$~meV, for which the system is in the $C=0$ phase (see Fig.~\ref{phasediagram}a), a different Hall sequence $\pm 4 \frac{e^2}{h}, -8 \frac{e^2}{h}$ is obtained, even though the system still has the same uniform magnetization $m_0$.  Instead, we believe that this effect is connected to the fundamentally different Hosftadter spectrum for the trivial and Chern bands, and hence, Wannier diagrams. 
The obtained results are summarized in Table~I. Wannier diagrams can be directly probed in experiment by performing Shubnikov–de Haas measurements on high-quality magic-angle TBG samples.\cite{Lu2019} We remark that the experimentally observed QH sequences are strongly sample-dependent,\footnote{The question of the Landau fan (LF) in TBG is a hard one. The original studies\cite{Cao2018} reported the Hall sequence, in units $e^2/h$: $\pm 4$, $\pm 8$, $\pm 12$ at neutrality, instead of expected $\pm 4$, $\pm 12$...\cite{Moon2012}, thus puzzling theorists. The consequent experiments\cite{Lu2019} on more homogeneous samples brought even further surprise with the QH sequence $\pm 2$, $\pm 4$, $\pm 8$. Moreover, LF with $\pm 2$, $\pm 4$ was reported in Ref.\cite{Sharpe2019};  LF with $\pm 2$, $\pm 4$, $\pm 6$ was reported in TBG stabilized by WSe$_2$;\cite{Arora2020} and LF with $- 2$, $- 4$ in the sample of Ref.\cite{Pierce2021}. For the high-field behavior of Landau fans, we refer readers to the recent observation of correlated Chern insulators in TBG e.g. in Refs. \cite{Pierce2021, Xie2021}}
 however the two leading contributions to the low-field Landau fans can be considered as $\pm 2 \frac{e^2}{h}$, $\pm 4 \frac{e^2}{h}$. It is certainly interesting that TBG with SkL in the $C=2$ phase (but not the $C=0$ phase) gives a reasonable Hall sequence, without demanding  $C_{3z}$ or $C_{2z}$ breaking.\cite{Zhang2019b}

\

\textbf{Conclusion.}---{In this Letter, we considered the influence of the real-space skyrmionic order on the dispersion and topology of  flat bands. For the particular example of magic-angle twisted bilayer graphene, we find that the commensurate skyrmionic order itself redefines both the band structure and topology of the system. Surprisingly, the real-space skyrmion order influences the electronic bands topology, giving the rise to an unexpected $C=2$ phase. To our knowledge, this is the first realistic system which provides a robust and affirmative answer to the question of the interplay between topologies in real and reciprocal spaces.\cite{Lux2021} According to the field-theoretical arguments,\cite{Sondhi1993,Abanov2001a,Hsu2013} the elementary excitations in this case are $Q=2 e$, a consequence important for the transport phenomena in twisted bilayer graphene. Experimentally, the reported effect manifests in  magnetotransport observables with the leading quantum Hall sequence $\pm 2 \frac{e^2}{h}$, $\pm 4 \frac{e^2}{h}$. This Landau fan fingerprint resonates well with experimental reports, and arises naturally in our model as a direct consequence of the emergent skyrmion order.

\newpage

\nonumber

{ 
 \footnotesize{ 

\noindent 
\textbf{METHODS} 

\

\noindent 
\textbf{Tight binding model for twisted bilayer graphene.}---The electronic structure calculations are performed with the tight-binding Hamiltonian capturing the  atomistic physics of twisted bilayer graphene in free-electron formalism,
\begin{align}
H=\sum_{ij} t_{ij} c^{\dag}_i c^{}_j, 
\end{align}
where  $c^{\dag}, c^{}$ are creation and annihilation operators and the numerical values of hopping terms $t_{ij}$ are computed within the  Slater-Koster formalism by taking into account hopping  of $p_z$ orbital on each atom.  
The hopping term between two $p_z$ orbitals is expressed as a sum of $\pi$ and $\sigma$ type overlapping,  $t_{ij}$$=$$V_i^\pi + V_{ij} ^\sigma$, conventionally approximated in the form\cite{Nam2017}
\begin{align}
t_{ij}=V_{0}^\pi\exp(-\frac{r-a_0}{r_0})\sin^2\theta+V_{0}^\sigma\exp(-\frac{r-d_0}{r_0})\cos^2\theta ,
\end{align}
where $r$$=$$|\vec r_{ij}|$ and $\theta$ is the angle between $\vec r_{ij}$ and the $\hat e_z$, $\theta$$=$$\angle \{ \vec r_{ij}, (0,0,1) \}$.  
The numerical parameters have been adapted from Ref.\cite{Nam2017}, hence $V_0^\pi$$=$$-2.7$ eV is the hopping integral between nearest-neighbor (NN) atoms in the same layer; $V_0^\sigma$=0.48 eV is the overlap between the aligned atoms in different layers;  here 
 $a_0$$=$$0.142$ nm and $d_0$=0.334 nm correspond to the nearest neighbor distance and the interlayer distance; 
$r_0$=0.184$a$ (we use $a$$=$$\sqrt{3} a_0$ as the lattice constant of graphene)   is the hopping strength decay length. 
The cutoff real-space is chosen at 9.84 \r{A}; in the outer region the coupling is negligible.
To the main approximation, the hopping amplitudes are barely affected by the real-space magnetic order. 
Hence on the tight-binding level the effect of local magnetization enters the Hamiltonian as on-site spin-orbit coupling terms on each atom.
In the basis $(\psi_i^\uparrow,\psi_i^\downarrow)$ the spin-orbit coupling reads as  $H_{\text{SOC}}$$=$$\boldsymbol{\sigma} \cdot \vec{m}(\vec r_i)$, where $\boldsymbol{\sigma}$$=$$(\sigma_x,\sigma_y,\sigma_z)$ are the conventional 2$\times$2 Pauli matrices.
At atomic position $i$, the on-site Hamiltonian term is
\begin{align}
 H_{i}^{\text{os}} & =\begin{pmatrix}1 & 0\\
0 & 1
\end{pmatrix}V_{i}
+\begin{pmatrix}0 & 1\\
1 & 0
\end{pmatrix} \mu_{x}
+\begin{pmatrix}0 & \text{-}i\\
i & 0
\end{pmatrix}\mu_{y}  
 +\begin{pmatrix}1 & 0\\
0 & \text{-}1
\end{pmatrix}[\mu_{z}+m_{0}] ,  
\end{align}
where we distinguish between the uniform magnetization ($m_0$$=$$\langle \vec m (\vec r) \rangle$) and the skyrmion order itself $\boldsymbol \mu$$=$$\vec m (\vec r)$$-$$\vec m_0$. The overall real-space magnetic phase is approximated by
\begin{align}
\vec m (\vec r)=\vec m_0 + \mu \sum_{j=1} (\vec S_{\vec q_j} e^{i \vec q_j \vec r} + \vec S_{-\vec q_j} e^{-i \vec q_j \vec r} ),
\end{align}
where for the skyrmion lattice it is sufficient to cut the Fourier terms by the first triade $\vec q_1 + \vec q_2 + \vec q_3$$=$$0$, with $|\vec q_i|$ commensurate with moir\'{e} periodicity. 
Adding such on-site terms on each orbital, the full spinful Hamiltonian with the underlying skyrmion order reads
\begin{align}
	H & =H_{\text{hop}}+H_{\text{os}} = \sum_{i\neq j}t_{ij}\sigma_{0}
	+ \sum_{i}\sigma_{0}V_{i}  \\
	& + \sum_{i}\sigma_{x} \mu_{x}(r_{i})+\sigma_{y}\mu_{y}(r_{i})+\sigma_{z}[\mu_{z}(r_{i})+m_{0}] .
\end{align}
Since the atomic relaxation has an effect on the interlayer coupling in AA and AB stacking region, and has dramatic effects on band flatness in TBG, we build the tight-binding hamiltonian based on the relaxed atomic structures. The atomic structure relaxation is done with classical potential on the LAMMPS package.\cite{PLIMPTON19951}
The classical force field contains the short range terms for covalent bonds and long-range terms which describes the Van der Waals interaction. 
The implemented long-range potentials are obtained from the parameterized version of the Kolmogorov-Crespi potential and confirmed to match the DFT results in Ref. \cite{Gargiulo_2017}. 
The short range interaction of long-range carbon bond order potential (LCBOP) was adopted from \cite{LCBOP_short} without modification.

\textbf{Hybrid Wannier functions formalism}---The Chern numbers of the moir\'e flat bands are calculated through  the formalism of hybrid Wannier function methods introduced by Soluyanov and Vanderbilt,\cite{Soluyanov2011} see also Refs.\cite{Taherinejad2014,WU2017} 
The Wannier functions $W_{n}(\vec {R})$ are conventionally defined as localized orbitals which are build through integral representation through the Bloch wave functions $u_k$,\cite{Wannier1962}  
\begin{align}
	W_{n}(\boldsymbol{R})=\frac{1}{(2\pi)^2} \int_{\text{BZ}}e^{i\boldsymbol{k}\cdot(\boldsymbol{r}-\boldsymbol{R})}|u_{n\boldsymbol{k}}\rangle .
\end{align}
On contrary, the hybrid Wannier functions (HWFs) are wannierized only along one of the two spatial dimensions (e.g. $x$), thus remaining a Bloch-periodic function in momentum along the other dimension (e.g. $k_y$). HWFs had proved to be useful to track the topological properties, in particular Chern numbers of flat electronic bands.\cite{Gresch2017} The underlying idea is to track the evolution of the hybrid Wannier functions $| W^{(h)}_{n} \rangle$ along a $k_y$-path, and the Chern number is given by the total winding of the Wannier centers.
 This approach is based on the observation that every nontrivial Chern number presents an obstruction for constructing maximally-localized Wannier functions.\cite{Marzari2012}
The numerical advantage we implement is: instead of finding the full 2D Wannier functions, we calculate the Wannier charge centers(WCCs) in one of the dimensions (e.g. $x$), and track it along the momentum on the other dimension($k_y$).
Here we give an example of the calculation of hybrid Wannier functions. 
For the 2D electronic system with 
its eigenstates in BZ  being $|u_{\vec k} (k_x,k_y) \rangle$ in  the conventional Bloch band framework, the Wannier charge centers (WCC) are defined through the phase accumulation, 
\begin{align}
 |u_{\vec k} (k_x+2\pi,k_y) \rangle= e^{i\phi_x} |u_{\vec k} (k_x,k_y) \rangle,
 \  \ \
\phi_x (k_y) \equiv  \text{WCC}(k_y). 
\end{align}
The numerical calculations are performed on the discretized BZ.
The phase factor $\phi_x$ is calculated by evolving from  $k_x=0$ to $k_x=2\pi$, and tracing the accumulated phase in each increment
\begin{align}
\phi_x (k_y) =\arg \left[ \prod_{i=1}^{n-1} |u_{\vec k}(k_x^i,k_y)\rangle \,  \langle u_{\vec k}(k_x^{i+1},k_y)| \right]
\end{align}
in order to preserve the smooth gauge. Thus defined, the WCC winding on the cylinder determines the Chern numbers by counting the number of times the WCC is crossing the boundaries.\cite{Gresch2017} 
The  winding of the Wannier center is giving the Chern number of a single or a composite band.
In multiple-band systems, the total Chern number is determined by the addition of the winding of all the bands. This method becomes particularly handy for evaluating the Chern numbers of the flat electronic bands.

\textbf{Effective continuum model}.---We  consider the twisted bilayer graphene tuned to the first magic angle, and coupled to the skyrmion magnetic order  with the same modulation vectors as the Moire pattern in TBG. 
At such small angles and large supercell scales, the effective physics of both the TBG and SkL can be described within the effective continuum models  (see e.g. Ref.\cite{TKV}). 
We first consider the SkL with  uniform (ferromagnetic) component to be  small. 
The effective Hamiltonian of the TBG coupled to the magnetic skyrmion Lattice $\vec m(\vec r)$ can be written as:
\begin{align}
\mathcal{H} = 
\begin{pmatrix}
- i v \boldsymbol{\sigma}_{-\theta/2} \nabla - \vec m (\vec r) \,  \vec s & T(\vec r) \\
T^{\dag}(\vec r) & - i v \boldsymbol{ \sigma}_{+\theta/2} \nabla - \vec m (\vec r) \vec s
\end{pmatrix}, 
\end{align}
where $\vec s$ is electron spin. Here $v$ is the input fermi velocity and rotated Pauli matrices for each layer defined as $\boldsymbol{\sigma}_{\theta}$$=$$e^{- i \theta \sigma_z /2} (\sigma_x, \sigma_y) e^{+ i \theta \sigma_z /2}$. The interlaying coupling is given by the 2x2 matrix
\begin{align}
T(\vec r) = \sum_{j=1}^{3}	T_j e^{- i \vec q_j \vec r},
\end{align}
where vectors $\vec q_1$, $\vec q_2$, $\vec q_3$ constitute the three-fold rotation with $\vec q_1+ \vec q_2 + \vec q_3$$=$$0$, $|\vec q_j|$$=$$k_{\theta}$$=$$2 k_D \sin (\theta/2)$, responsible for the Moire lattice formation. We here use $\vec q_1$$=$$k_{\theta} (0, -1)$, $\vec q_{2,3}$$=$$k_{\theta} (\pm \sqrt3/2, 1/2)$. In general the symmetries of the TBG allow coupling matrices $T_j$ of the form,
\begin{align}
 T_{j+1} = w_{\text{AA}}  \, \sigma_0 + w_{\text{AB}} \, (\sigma_x \cos n \phi + \sigma_y \sin n \phi ) ,
\ \ \ \phi = \pi/3
\nonumber
\end{align}
We consider a magnetic skyrmion Latice (SkL) matched to the twisted graphene bilayer in the way that both the supercell periodicities coincide precisely, and thus the skyrmion lattice is build on the $\vec Q_1$$=$$\vec b_1$, $\vec Q_2$$=$$- \vec b_2$,  $\vec Q_3$$=$$ \vec b_3$  commensurate with the Moire lattice. 
The skyrmion lattice order parameter to a good accuracy can be written as 
\begin{align}
\vec m (\vec r) = \vec m_0 + \mu \sum_{j=1} (\vec S_{\vec b_j} e^{i \vec b_j \vec r} + \vec S_{-\vec b_j} e^{-i \vec b_j \vec r} ),
\end{align}
where we are primarily building the skyrmion lattice on the moire modulation vectors $\vec b_1$$=$$\vec q_2 - \vec q_1$, $\vec b_2$$=$$\vec q_3 - \vec q_1$, $\vec b_3$$=$$\vec q_3 - \vec q_2$,  (thus $\vec Q_1 + \vec Q_2 + \vec Q_3$$=$$0$ form the SkL crystalline), plus further momenta terms if needed. 
 Here $\vec m_0 || \vec B$ is a uniform component proportional to the strength of external magnetic field. The typical band structures are listed in Supplementary Material, and qualitatively coincide with the tight binding calculations. Importatly, the continuum model shows that the bands cannot be made perfectly flat once the SkL is added.

\textbf{Hofstadter diagrams: Lanczos algorithm.}---
For computation of the Hofstadter diagrams the diagonalization is performed on a magnetic supercell  retaining the periodicity of the boundary conditions. 
Following the Landau gauge, $\vec{A}$$=$$(-By,0,0)$, each  tight-binding calculation on such magnetic supercells contain diagonalization of large matrices, usually up to dimension $\sim$ $10^7$. 
Regular diagonalization methods(with complexity $O(N^3)$ fail quickly in such a large job due to the consumption of time and memory.
In order to perform the required calculation of the Hofstadter and Wannier diagrams, we make use of the Lanczos algorithm implemented in the WannierTools package.\cite{WU2017} 
The Lanczos algorithm solves the eigenvalue or density of states problems by iteration, requiring much less memory than the direct diagonalization.         
The idea of the algorithm is to iteratively apply the Hamiltonian matrix to the set of initial vectors, and subsequently transform the Hamiltonian  transformed into a tridiagonal form during the process.
To diagonalize a Hamiltonian $H$, the algorithm starts with an initial vector $|v_1\rangle$ whose norm is 1,  each iteration is done as following: 

\noindent
	1. The main diagonal element is given by $\alpha_i=\langle v_i |H |v_i \rangle $.

	\noindent
	2.   Get the second vector $|v_i+1\rangle=H|v_i\rangle - \alpha_i|v_i \rangle$.
	
	\noindent
	3.  Find the sub-leading diagonal  $\beta_i=|| (|H v_i+1 \rangle) ||$.
	
	\noindent
	4.   Rescale $|v_{i+1} \rangle =|v_{i+1}\rangle  /\beta_i$. 
 
 	\noindent
The operations are further performed iteratively until the convergence is achieved.
By the above iteration, the Hamiltonian is transformed  into a tridiagonal form:
\begin{align}
H_{\text{tri}} = \begin{pmatrix}
	\alpha_{1} & \beta_{1} & 0 & 0 & 0\\
	\beta_{1} & \alpha_{2} & \beta_{2} & 0 & 0\\
	0 & \beta_{2} & ... & \beta_{n-1} & 0\\
	0 & 0 & \beta_{n-1} & \alpha_{n} & \beta_n\\
	0 & 0 & 0 & \beta_n & ...
\end{pmatrix} .
\end{align}
Having the tridiagonalized Hamiltonian, the density of states is thereafter evaluated through the Green's functions,
\begin{align}
\rho(E+i\varepsilon)=-\Im \Tr([G(E+i\varepsilon)])=-\Im \Tr[(E+i\varepsilon)I-H]^{-1} .
\label{DOS-L}
\end{align}
One further  applies the continued fraction method to calculate $G$$=$$[(E+i\varepsilon)I-H]^{-1}$.
For the complex frequency $\Omega$$=$$E+ i \varepsilon$, the continued fraction representation is
\begin{align}
G(\Omega)  = (\Omega -H_{\text{tr}})^{-1}
 =\frac{1}{\Omega-\alpha_{1}-\frac{\beta_{1}^{2}}{\Omega-\alpha_{2}-\frac{\beta_{2}^{2}}{\Omega -...}}} .
 \label{Green-L}
\end{align}
Further-on, the DOS in the selected energy range is obtained by evaluating the Green's function iteratively.
}}

\

\

\textbf{Acknowledgments}. The authors thank Bertrand Halperin and Henrik Rønnow for useful discussion. A.K. thanks Ashvin Vishwanath for multiple discussions on the topic. The project was supported by the Branco Weiss Society in Science, ETH Zurich, through the research grant on flat bands, strong interactions and SYK physics, and by the Swiss National Science Foundation, grants No. 172543 and P2ELP2{\_}175278. Computations have been performed at the Swiss National Supercomputing Centre (CSCS) under project s1008 and the facilities of Scientific IT and Application Support Center of EPFL.

\bibliography{Refs}

\end{document}